\documentclass[iop,revtex4]{emulateapj}

\usepackage{natbib}
\usepackage{graphicx}  
\usepackage{lscape}
\usepackage{hyperref}

\newcommand{\galex}{{\it GALEX}}
\newcommand{\kepler}{{\it Kepler}}

\slugcomment{Submitted to the Astrophysical Journal}
\shorttitle{GALEX-Kepler Catalog}
\shortauthors{Olmedo et al.}

\begin{document} 

\title{Deep GALEX UV Survey of the Kepler Field I: 
Point Source Catalog}

\author{
Manuel Olmedo\altaffilmark{1,2},
James Lloyd\altaffilmark{3,4},
Eric E. Mamajek \altaffilmark{2},
Miguel Ch\'avez\altaffilmark{1},
Emanuele Bertone\altaffilmark{1},
D. Christopher Martin\altaffilmark{5},
James D. Neill\altaffilmark{5}
}

\altaffiltext{1}{Instituto Nacional de Astrof\'{i}sica Optica y 
Electr\'{o}nica Luis Enrique Erro \#1, CP 72840, Tonantzintla, Puebla, Mexico}
\altaffiltext{2}{University of Rochester, Department of Physics \& Astronomy, Rochester, NY, 14627-0171, USA} 
\altaffiltext{3}{Department of Astronomy, Cornell University, Ithaca, NY, USA}
\altaffiltext{4}{Carl Sagan Institute and Cornell Center for Astrophysics and Planetary Science, Cornell University, Ithaca, NY, USA}
\altaffiltext{5}{California Institute of Technology, 1200 East California Boulevard, MC 278-17, Pasadena, CA 91125, USA} 

\email{olmedo@inaoep.mx}

\begin{abstract} 
We report observations of a deep near-ultraviolet (NUV) survey of the
\kepler\ field made in 2012 with the Galaxy Evolution Explorer (GALEX)
Complete All-Sky UV Survey Extension (CAUSE).
The GALEX-CAUSE Kepler survey (GCK) covers 104 square degrees of the
\kepler\ field and reaches limiting magnitude NUV $\simeq$ 22.6 at
3$\sigma$.
Analysis of the GCK survey has yielded a catalog of 669,928 NUV
sources, of which 475,164 are cross-matched with stars in the
\kepler\ Input Catalog (KIC).
Approximately 327 of 451 confirmed exoplanet host stars and 2614 of
4696 candidate exoplanet host stars identified by \kepler\ have NUV
photometry in the GCK survey.
The GCK catalog should enable the identification and characterization
of UV-excess stars in the \kepler\ field (young solar-type and
low-mass stars, chromospherically active binaries, white dwarfs,
horizontal branch stars, etc.), and elucidation of various
astrophysics problems related to the stars and planetary systems in
the \kepler\ field.
\end{abstract}


\section{Introduction}

The fast growth of exoplanet detections has motivated the derivation
of more accurate fundamental stellar properties (e.g. mass, radius,
age, etc.) and the connection between these properties and those of
their evolving planetary systems.
The precision with which the exoplanets' parameters can be estimated
directly depends on the precision associated to the properties of
their stellar hosts.
%
%
Of particular importance is the stellar age, since one of the major
goals in the study of exoplanetary systems is to establish their
evolutionary stage, and how this compares to the properties of our own
solar system.


Reliably age-dating solar-type field stars is notoriously difficult.
For these stars, alternative methods to isochrone fitting techniques
have been explored.
Chromospheric activity and stellar rotation are among the more
reliable observables for stellar age estimation of Sun-like main
sequence stars.
The most common and accessible proxy for stellar activity has been the
emission in the Ca\,II resonance line in the optical-ultraviolet
spectral interval \citep[e.g.][]{Mamajek08}.
Alternative proxies have also been identified and include the high
contrast emission of Mg\,II line in the ultraviolet (UV) and the
continuum UV excess \citep[e.g.][]{Findeisen11,Olmedo13}.
For these stars, ultraviolet radiation originates in the hot plasma of
the upper stellar atmospheres at temperatures of $\sim
10^{4}-10^{6}$~K, heated by non-thermal mechanisms, such as acoustic
and magnetic waves, generated by convection and rotation
\citep[e.g.][]{Narain96,Ulmschneider03}.
As the star ages, it loses angular momentum due to magnetic braking
\citep{Mestel68,Kawaler88}, slowing the rotation and affecting the
stellar dynamo, which in turn decreases the magnetic field leading to a
decrement of the UV emission.
In this first paper of a series aimed at investigating the UV
properties of stars, we present a complete catalog of UV sources
detected by the Galaxy Evolution Explorer
\citep[\galex;][]{Martin05,Bianchi14} in the field of the
\kepler\ Mission \citep{Basri05}.


The \kepler\ field (i.e., the 100 square degree field of the original
\kepler\ mission) has been fully surveyed at different optical and
infrared bandpasses \citep{Lawrence07,Brown11,Everett12,Greiss12}.
Its stellar content, mainly comprised of field stars, with over 450
confirmed\footnote{http://kepler.nasa.gov/ as of 11 September 2015}
exoplanet host stars, has rapidly become one of the most studied
stellar samples and regions of the sky.
The \kepler\, field and associated survey data comprise a potentially
valuable resource for studying the age-activity-rotation relation for
low-mass stars.
Rotation periods (and ages) of stars in the \kepler\ field have been
determined in a number of studies
\citep[e.g.][]{Reinhold13,Walkowicz13, Garcia14, McQuillan14,
  Nascimento14, Meibom15}.


The \galex\ UV space observatory was launched in 2003 and was operated by
NASA until 2011 \citep{Martin05,Bianchi14}.
Afterwards, the \galex\ mission was managed by Caltech for about a
year as a private space observatory, until the satellite was turned
off in June 2013.
\galex\ is a 50 cm Ritchey-Chr\'etien telescope with a 1$^{\circ}$.25
wide field-of-view, equiped with a near-UV (NUV; 1771-2831~\AA,
$\lambda_{\rm eff}$ = 2271~\AA) and a far-UV (FUV; 1344-1786~\AA,
$\lambda_{\rm eff}$ = 1528~\AA) detectors.
During it's main mission, \galex\ carried out the All-Sky, Medium, and
Deep Imaging Surveys, with exposure times of $\sim$100, $\sim$1000,
and $>$10000 sec, respectively, measuring more than 200 million
sources \citep{Bianchi14}.
The continuation phase, called the GALEX Complete All-Sky UV Survey
Extension (CAUSE), was funded by several consortia, with the main goal
of extending observations in the NUV band to the Galactic plane, which
was only scarcely mapped during the main mission, due to restrictions
on the maximum target brightness.\\


In this work, we describe the creation of a deep NUV photometric
catalog with nearly full coverage of the \kepler\ field using CAUSE
observations.
In Section \ref{sec:obs} we introduce the observations, and
characteristics of the data.
In Section \ref{sec:extr} we explain the procedure for extracting the
NUV point sources.
Section \ref{sec:cat} describes the point source catalog, and its
crossmatch with the \kepler\ Input Catalog (KIC) and the
\kepler\ Objects of Interest (KOI) catalog is presented in Section
\ref{sec:cross}.
Finally, the publicly available GALEX-CAUSE (GCK) catalog of NUV
sources is described in Section \ref{sec:file}.



\section{Observations}
\label{sec:obs}

As part of the CAUSE survey, Cornell University funded 300 orbits to
complete the spatial coverage of the \kepler\ field through
August-September 2012 (PI J. Lloyd).
The dataset of \galex\ CAUSE Kepler field observations (hereafter GCK)
is composed of 180 tiles that cover the \kepler\ field
\citep{Borucki03, Basri05, Brown05, Latham05}; each tile has 20 visits
on average.
These observations sample timescales from a millisecond to a month,
and can be used to identify variable targets and exotic sources on
such timescales.
The GCK dataset provides spatial coverage of the \kepler\ field in the
\galex\ NUV band.


The standard \galex\ pointed observation mode adopted for the surveys
during the primary NASA \galex\, mission employed a 1\arcmin.5 spiral
dither pattern.
This dither moves the sources with respect to detector artifacts and
prevents a bright source to saturate one position on the detector, which is
subject to failure if overloaded.
The \galex\ data pipeline processes the photon arrival times and
positions with an attitude solution that reconstructs an image of the
sky for a single tile, 1$^{\circ}$.2 in diameter around the pointing
center.

\begin{figure}[h!]
\epsscale{1.1}
\plotone{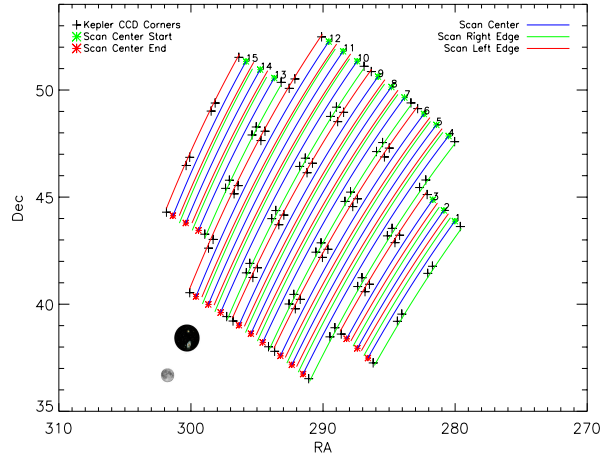}
\caption{Scan mode observations of GCK survey covering the
  \kepler\ field.  The field of view of \galex\ and a full moon are
  shown for comparison.  The numbers on the upper right edges
  correspond to the scan numbers.  The plus symbols corresponds to the
  edges of \kepler\ detectors.
\label{fig:scanmode}}
\end{figure}

In 2012, a drift mode was adopted, which scanned a strip of the sky
along a great circle. 
For GCK observations, these scans run as long as 12$^{\circ}$.
The scan mode processing uses a pipeline adapted from the pointed mode
observations\footnote{http://galex.stsci.edu/GR6/?page=scanmode}.
To adapt the scans to the standard tile processing pipeline, they are
processed in tile sized images resulting in 9 and 14 images for the
short scans (1$-$3 and 13$-$15) and long scans (4$-$12), respectively
(shown in Fig.~\ref{fig:scanmode}).
Thus for the 180 tiles of the GCK data, with each visited an average
of 20 times, we have a total of $\sim$ 3200 images.


\begin{figure*}[h!]
\epsscale{1.1}
\plotone{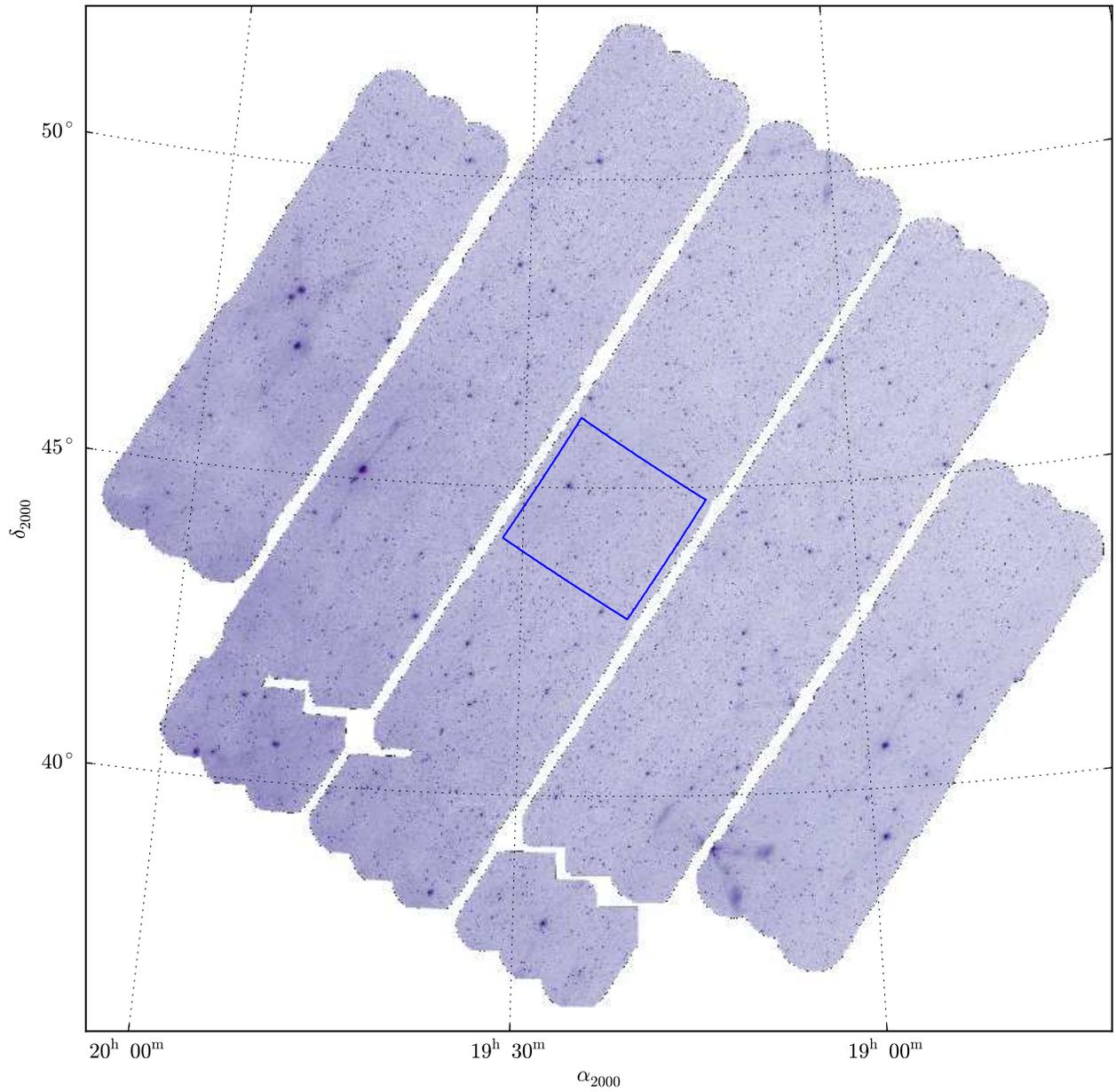}
\caption{Intensity image mosaic of \galex\ CAUSE NUV observations. The
  blue square corresponds to the central CCD of the
  \kepler\ telescope.
  \label{fig:intensity_mosaic}}
\end{figure*}


\begin{deluxetable}{lll} 
\tabletypesize{\scriptsize}
\tablecaption{Files of GCK dataset}
\tablewidth{0pt}
\tablehead{
\colhead{Fits name} & Image type & Units}
\startdata
nd-count.fits & count image            & photons/pix \\
nd-rrhr.fits  & effective exposure map & seconds/pix \\
nd-int.fits   & intensity map          & photons/second/pix \\
nd-flags.fits & artifact flags image   & .. \\
xd-mcat.fits  & catalog of sources     & .. \\
\enddata
 \label{table:products}
\end{deluxetable}


The GCK data were processed with the \galex\ scan-mode pipeline and
subsequently delivered to Cornell in the form of packs of 5 images
(see Table~\ref{table:products} for details) per visit for each tile.
For details on these files or any technical information concerning the
\galex\ mission see: {\small
  http://www.galex.caltech.edu/wiki/Public:Documentation.}

\begin{figure}[h!]
\epsscale{0.8}
\plotone{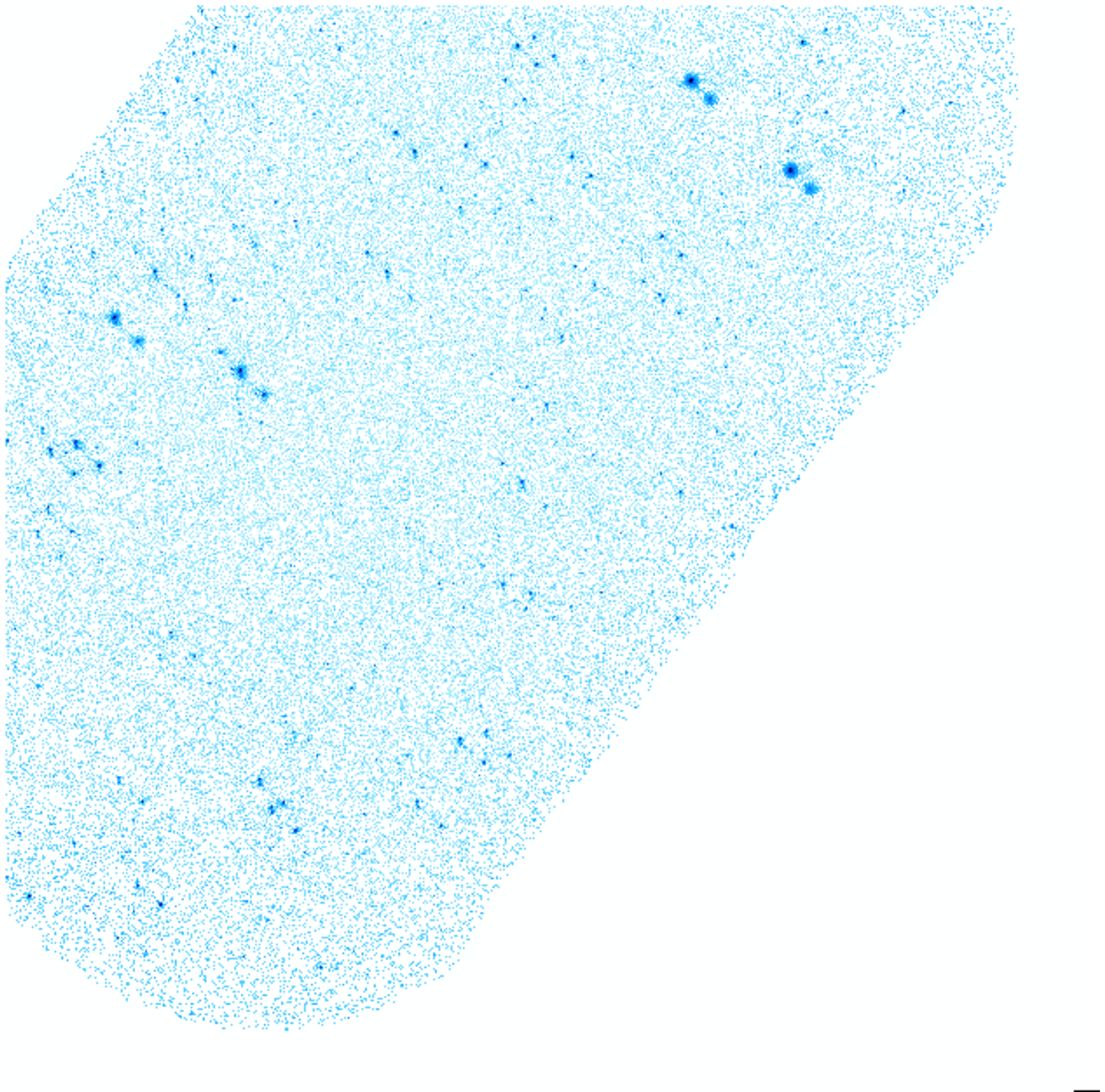}
\caption{\galex\ image corresponding to scan 1, image 2, visit 2, as
  an example of a failure of the pipeline processing: bright sources
  appear as double objects. \label{fig:streaked}}
\end{figure}

A mosaic of the \galex\ CAUSE NUV observations is shown on
Fig.~\ref{fig:intensity_mosaic}.  The superimposed blue box
corresponds to the central CCD of the \kepler\ detector array.  Most
of the data exhibit the exquisite photometric and astrometric
stability and reproducibility expected for a space
observatory. However, the \galex\ pipeline failed to correctly process
a small fraction of the images.  A thorough visual inspection showed
that there are about 450 images affected by doubling or ghosting of
sources. An example of this effect is illustrated in
Fig.~\ref{fig:streaked}.  All these images were excluded from our
analysis.

\begin{figure}[h!]
\epsscale{1.1}
\plotone{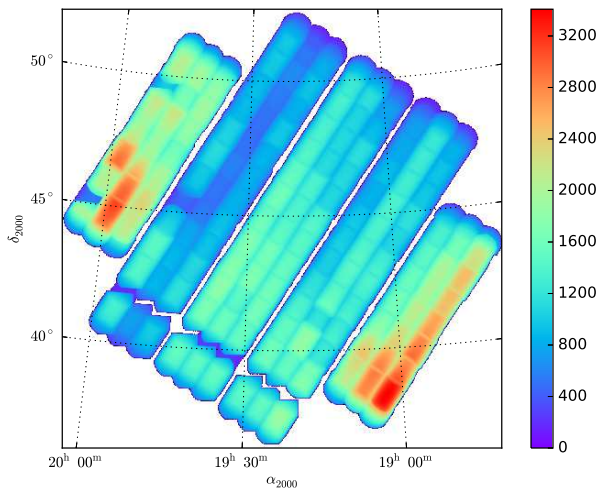}
\caption{Mosaic of the effective exposure time of the \galex\ CAUSE
  observations.
 \label{fig:exposure_mosaic}}
\end{figure}

\begin{figure}[h!]
\epsscale{1.1}
\plotone{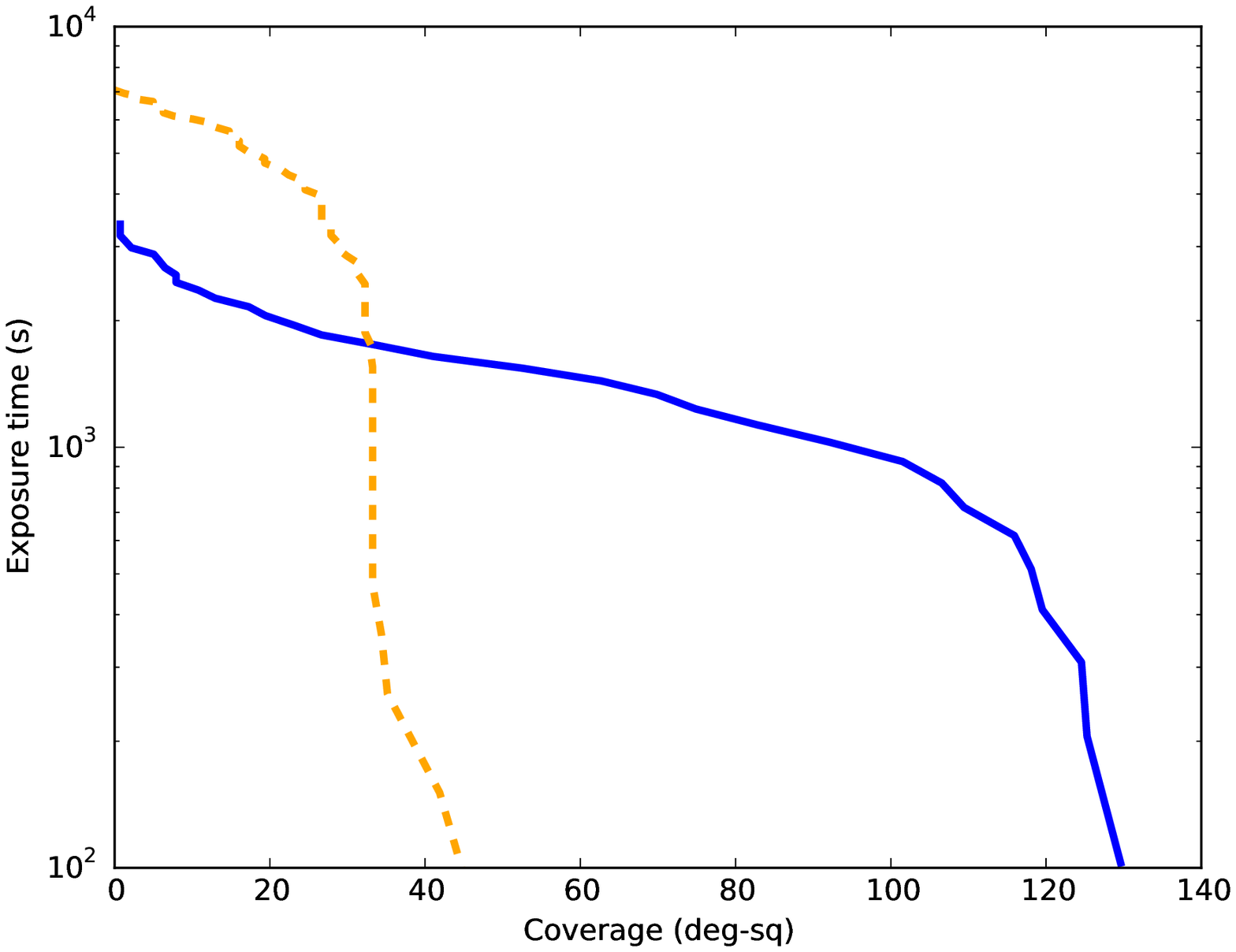}
\caption{Cumulative distributions of exposure time of the sky coverage
  of the GCK \kepler\ field (solid blue line). Observations from the
  \galex\ release 6 (GR6) are represented with the dashed orange line.
 \label{fig:exptime_distribution}}
\end{figure}

The final exposure time map is presented in
Fig.~\ref{fig:exposure_mosaic}.
 Some tiles lack a single good visit and are located at the end of a
 scan, when the spacecraft was executing a maneuver.  Because of this,
 the 13th image of scans 4, 5, 6, 9 and 10 were not included in the
 catalog construction (leaving a total of 175 tiles), representing a
 loss of no more than 0.5\% coverage of the \kepler\ field.  Within
 the main \galex\, surveys, fractions of the \kepler\ field were
 partially surveyed \citep{Smith11}. A cumulative distribution of the
 exposure time in function of sky coverage is shown in
 Fig.~\ref{fig:exptime_distribution} (solid blue line), and the orange
 dashed line corresponds to the \galex\ release 6 (GR6).


\section{Assembly of a Catalogue of NUV Sources}
\label{sec:extr}

The construction of the GCK catalog can be summarised in 4 stages,
each of which is discussed in this section. In the first stage, "Image
Co-adding'', the available single epoch visits for each tile are
co-added. Next, the "Background Estimation'' is carried out from the
intensity image using a modified $\sigma$-$\kappa$ clipping
method. The "Source Extraction and Photometry'' uses the software
SExtractor \citep{Bertin96} to first detect and then perform
photometry on the background-subtracted intensity image for each tile
obtained in the previous stage.  In the final stage the catalogs from
each of the 175 tiles are combined, and duplicate objects, low S/N
sources, and other possible spurious sources are carefully removed.

\subsection{Image Co-adding}

Prior to co-adding all epochs for a given tile, each image was
visually inspected, discarding visits presenting the source doubling
or ghosting issue (Fig. \ref{fig:streaked}). Since the images already
have an astrometric solution, and furthermore, each epoch for a given
tile are aligned, co-adding only requires an arithmetic sum.  For each
tile, the process was the following:

\begin{enumerate}
\item Construction of the co-added count image as the arithmetic sum
  of individual count intensity images ({\it nd-count.fits}).
\item Construction of the co-added effective exposure image as the
  arithmetic sum of individual effective exposure images
  ({\it nd-rrhr.fits}). 
\item Construction of a combined flag image as the logic OR of the
  individual flag images ({\it nd-flags.fits}).
\item Calculation of the the ratio of the co-added count and effective
  exposure images to obtain the final intensity image.
\end{enumerate}

\subsection{Background and Threshold Estimations}

The source extraction and photometry methodology follows that of the
\galex\ pipeline \cite[reported
  in][]{Morrissey07}\footnote{\scriptsize{http://www.galex.caltech.edu/wiki/Public:Documentation/Chapter\_104}}.
Prior to this step, background and threshold images are built, as
required by SExtractor.

\textit{Background estimation}: the construction of a background image
consists of an iterative $\sigma$-$\kappa$ clipping method. The
\textit{count} image is divided into square bins 128~pix wide. In each
bin, the local background histogram is built using the Poisson
distribution (due to the low NUV background count rates), and the
probability ${P_k}(x)$ of observing $k$ events for a mean rate $x$ is
calculated. Pixels with a ${P_k}(x) < 1.35 \times {10^{ - 3}}$
(equivalent to a $3\sigma$ level) are iteratively clipped until
convergence is reached.  Then a $5 \times 5$~pix median filter is
applied to decrease the bias by bright sources. The bin mesh is
upsampled to the original resolution and divided by the effective
exposure image to produce the final \textit{background} image, which
is subtracted from the \textit{intensity} image to produce the
\textit{background-subtracted intensity} image. 

\textit{Weight threshold image}: this image provides the threshold for
potential detections. The \textit{count} image is again divided in
128$\times$128~pix bins. In each bin the value of $k$, in counts/pix,
which corresponds to probability level of $3\sigma$ is computed and
stored to produce a threshold map. This map is upsampled to the
original resolution and divided by the effective exposure image, in
order to obtain a threshold image in counts/sec/pix.  The final step
is to compute a \textit{weight threshold} image by dividing the
\textit{background-subtracted intensity} image by this last threshold
image.

\subsection{Source Extraction and Photometry}

The detection and photometry process is carried out with SExtractor
working in dual mode: the \textit{weight threshold} image is used for
detecting sources, while their photometry is computed on the
\textit{background-subtracted intensity} image.
The SExtractor parameters \textit{THRESH\_TYPE} and
\textit{DETECT\_THRESH} are set to "absolute" and "1"; in this way
SExtractor will consider all pixels with values above 1 in the
\textit{weight threshold} image as possible detections.
SExtractor is executed for each of the 175 tiles in the GCK data,
delivering detection and photometry of each source.
The photometric error $dm$ in magnitude is calculated following the
\galex\ pipeline\footnote{Section 4 from: http://galexgi.gsfc.nasa.gov/docs/galex/Documents /GALEXPipelineDataGuide.pdf }:

\begin{equation}
\label{eqn:eqlabel}
df = {{\sqrt {(f + s\Omega )t} } \over t} , \; \; \; \;
dm = 1.086 \cdot df/f,
\end{equation}

where $f$ is the flux from the source in counts/sec, $s$ is the sky
level in counts/sec/pix, $\Omega$ is the area over which the flux is
measured and $t$ is the effective exposure time in seconds.

\subsection{Artifact Identification}

\begin{figure}[h!]
\epsscale{1.2}
\plotone{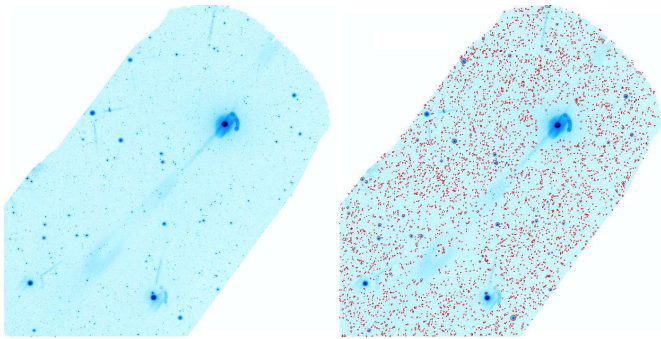}
\caption{{\it Left panel}: Co-added intensity image for image 6 of scan 2. 
{\it Right panel}: the same image with source detected marked as
  red points. Note the absence of detection in the area affected by instrumental artifacts. \label{fig:artifacts}}
\end{figure}

In the \galex\ imagery, there are various artifacts, not all of which
are automatically detected by the \galex\ pipeline.
The worrisome non-flagged artifacts are large diffuse reflections
within the field surrounding very bright stars.
The shapes of these artifacts have quite different morphologies,
however the most common shapes are long thin cones, halos, and
horseshoe-shaped extended reflections. An example is shown in
Fig. \ref{fig:artifacts}.  These artifacts bias the background and
affect source detection, mainly producing false positives.


\subsubsection{False Positives}

In order to remove extended objects and spurious detections, caused by
non-flagged artifacts in the images, we design suitable criteria to
remove them, while at the same time minimizing the loss of genuine
sources.  The criteria were defined considering the geometric
characteristics of the aperture fitted by SExtractor to the flux
profile of detection and the signal-to-noise ratio (S/N):

\begin{itemize}
\item Semi-minor axis $>$ 60\arcsec
\item Eccentricity $>$ 0.95
\item S/N $<$ 1.05
\item S/N $<$ 1.5 \& Semi-major axis $>$ 10\arcsec.
\end{itemize}

The first two criteria are intended to remove large and/or extended
sources, including false detections inside extended artifacts. The
third criteria discards too low S/N detections, while the purpose of
the fourth criteria is to remove detections at the border of images,
where the flux inside the SExtractor apertures may not be reliable.
We defined the above thresholds through a trial and error process,
aimed at discarding most of the false positives, while minimizing the
loss of real sourcesones.  The right panel of Fig. \ref{fig:artifacts}
shows the source detections after removing false positives.

\subsubsection{Artifact Flags}
 
The \galex\ pipeline produces a \textit{flag} image for each
tile. This image marks the pixels where the intensity image is
affected by some artifacts or where artifacts were removed.  Each
detected source has the {\it artifact\_flags} keyword, that is a
logical OR of the artifact flags for pixels that were used to compute
its photometry. These flags are summarized in Table
\ref{table:artifactflags}. The flags were developed
for the dither mode observations and in some cases are not
directly applicable to the scan mode observations. For
example, in dither mode observations, only the outer region of
any field is close to the detector edge, but in scan mode the
detector edge crosses nearly the entire field for a fraction
(but only a fraction) of the integration. The detector edge
proximity flag (Flag 6) is therefore set by the pipeline,
but it does not induce errors in the data quality.
  Flags 1 and 5 indicate the possible presence of reflections near 
 the edge of the FOV. However, we have found that in many instances
 the regions affected by this latter artifact do not actually show any obvious
problem. We therefore consider Flag 1 and 5 as non influential.
Detections with flags 2, 3, and 10 have been discarded because their photometry 
is affected by an unreliable background level determination. An example of an image
which presents regions marked by the flag 3, the dichroic reflection,
is shown in Fig.~\ref{fig:artifacts}, where one can distinguish
arc-shaped features around the three brightest stars in the image.

The GALEX pipeline flags were developed for the dither mode observations and in some cases are not directly applicable to the scan mode observations in the same way as the dither mode observations.  For example, in dither mode only the outer region of any field is close to the detector edge, but in scan mode the detector edge crosses nearly the entire field for a fraction (but only a fraction) of the integration.  The detector edge proximity flag is therefore set by the pipeline, but it does not induce errors in the data quality


\begin{deluxetable}{lll}[h!] 
\tabletypesize{\scriptsize}
\tablecaption{Artifact Flags}
\tablewidth{0pt}
\tablehead{Number & Short name & Description}
\startdata
1(1)   & edge        & Detector bevel edge reflection (NUV only) \\
2(2)   & window      & Detector window reflection (NUV only)\\
3(4)   & dichroic    & Dichroic reflection\\
4(8)   & varpix      & Variable pixel based on time slices\\
5(16)  & brtedge     & Bright star near field edge (NUV only)\\
6(32)  & detector rim& Proximity($>$0.6 degrees from field center)\\
7(64)  & dimask      & Dichroic reflection artifact mask flag\\
8(128) & varmask     & Masked pixel determined by varpix\\
9(256) & hotmask     & Detector hot spots\\
10(512)& yaghost     & Possible ghost image from YA slope \\
\enddata
\tablecomments{From the GALEX documentation, chapter 8 
(http://www.galex.caltech.edu/wiki/Public:Documentation/Chapter\_8).
Artifacts 7, 8, 9 do not apply to the the GCK catalog.}
 \label{table:artifactflags}
\end{deluxetable}


\section{The GCK UV source catalog}
\label{sec:cat}


After processing each file through false positive removal and
flagging, the 175 catalogs are combined to produce a single point
source catalog for the whole GCK field.  In the case of sources with
multiple detections (due to small overlap between tiles), the
measurement with the highest S/N was retained.

\begin{figure}[h!]
\epsscale{1.2}
\plotone{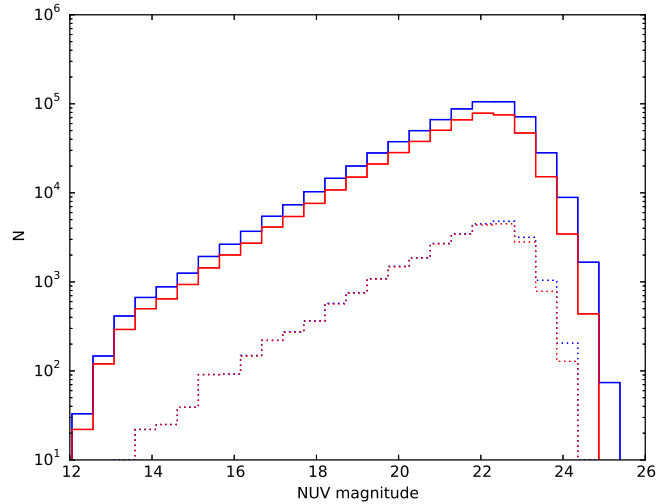}
\caption{Distribution of NUV detections. Blue line is the distribution
  of objects on the GCK catalog, red line are the matched objects with
  the KIC. Dotted lines correspond to a sample of detections inside a
  the sky region defined on Figure ~\ref{fig:intensity_mosaic}.
 \label{fig:distnuv}}
\end{figure}
\begin{figure}[h!]
\epsscale{1.2}
\plotone{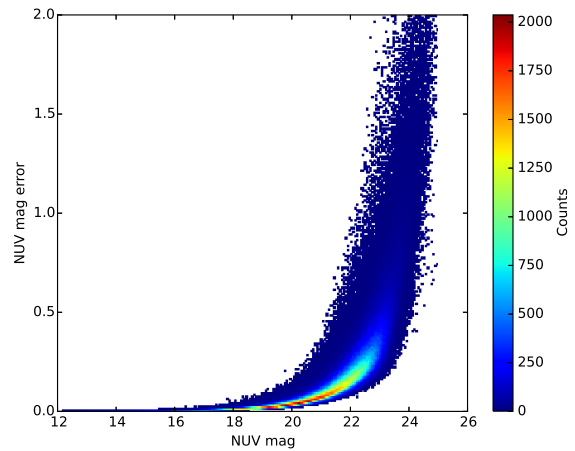}
\caption{NUV photometric error distribution for the GCK catalog. The
  color scale indicates the density of the sources in the
  plot. \label{fig:nuverror}}
\end{figure}

The resulting GCK catalog contains 660,928 NUV sources.  The NUV
brightness distribution of the GCK sources is shown in
Fig.~\ref{fig:distnuv} (blue lines), while the photometric error is
plotted in Fig.~\ref{fig:nuverror}. One can see that a typical error
for sources with NUV$<$22.6~mag is less than 0.3~mag.


 The GCK Catalog includes a few thousand objects with
  NUV$\lesssim$15.5, whose flux estimation is affected by the
  non-linearity of the GALEX NUV detector \citep[discussed
    by][]{Morrissey07}. The magnitude of this non-linearity effect can
  be as large as $\sim$1.0 mag at NUV\,$\simeq$\,12~mag.  In a recent
  analysis of the GALEX absolute calibration, \cite{Camarota14}
  identified a non-linear correlation when comparing GALEX data of an
  extend sample of white dwarf stars with predicted magnitudes from
  model atmospheres and spectroscopic data collected by the
  International Ultraviolet Explorer (IUE). They provide a correction
  that properly converts the GALEX NUV magnitudes to the Hubble Space
  Telescope photometric scale.  We applied this correction, based on
  the data in Table 3\footnote{Note that the correct C0 coefficient is 14.0821 for the IUE synthetic
fluxes (Camarota, priv. comm.)} of \cite{Camarota14}, to all GCK stars with
  NUV$\le$15.71~mag, for which we also provide the original GALEX
  magnitudes. We also obtain an uncertainty of 0.22~mag, associated to
  the magnitude conversion, by computing the standard deviation of the
  \cite{Camarota14} data with respect to their best fit.


\begin{figure}[h!]
\epsscale{1.2}
\plotone{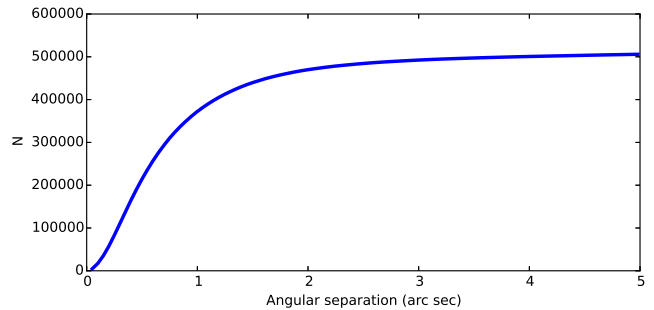}
\caption{Cumulative distributions of separations of the GCK - KIC cross-match.  
 \label{fig:separations}}
\end{figure}


\section{Crossmatch with KIC and KOI}
\label{sec:cross}

The positions of the GCK objects were cross-matched with the KIC \citep{Brown11}, using
a 2\arcsec.5 search radius. This value is compatible with the
astrometric precision of \galex\ observations, extending the crossmatch beyond 
this radius do not increase the number of matches more than 1\%, as can be seen 
in the Fig.~\ref{fig:separations}. 
 The cross-match resulted in 475,164 GCK objects with KIC counterparts. 
 We would like to remark that a smaller search radius would significantly decrease
the number of detections, while a larger radius would only increase
the KIC sources to be associated with a single NUV source. In the
final catalogue we provide the identification of KIC counterparts.

\begin{figure}[h!]
\epsscale{1.2}
\plotone{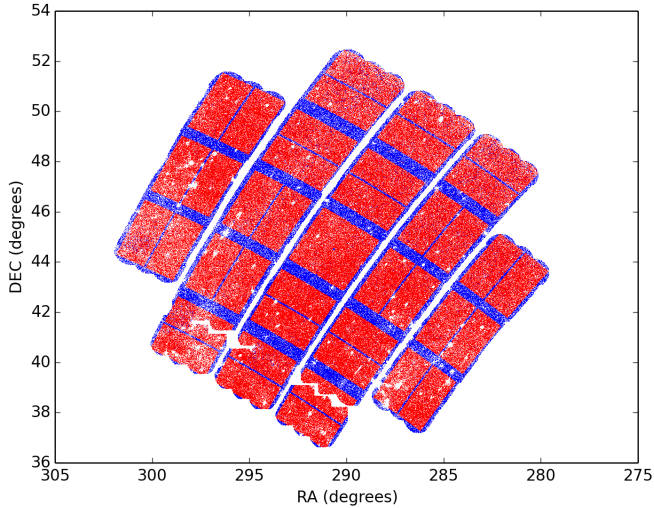}
\caption{Spatial distribution of the GCK sources in the
  \kepler\ field. Red dots indicate GCK sources with a counterpart in
  the KIC, while blue dots mark GKC sources without a KIC match. The
  saw-like regions in the south-east show the tile that were lost due
  to the pipeline failure (see Sect.~\ref{sec:obs}). Small circular
  empty regions are due to very bright stars, whose surrounding
  artifacts compromised nearby sources detections. \label{fig:detfov}}
\end{figure}

The spatial location of these objects are plotted as red points in
Fig.~\ref{fig:detfov}, where they obviously coincide with the position
of the \kepler\ satellite detectors. Their NUV brightness distribution
is also shown in Fig.~\ref{fig:distnuv}.  The constant ratio
difference (up to NUV$\sim$22.5~mag) between GCK and KIC distributions
seen in Fig. \ref{fig:distnuv} is due to incomplete coverage of the
\kepler\ field (caused by the gaps between the \kepler\ mission
detectors (Fig.~\ref{fig:detfov}).  In order to avoid this issue, we
also show in Fig. \ref{fig:distnuv} the number distribution of GCK
sources and their KIC counterparts of the objects on the central
\kepler\ detector(see Fig.~\ref{fig:intensity_mosaic}); we can see
that almost all GCK sources have KIC counterparts up to
NUV$\sim$22.5~mag.


We also cross-matched the GCK catalog with the KOI catalog available in MAST 
through the tool CasJobs\footnote{http://mastweb.stsci.edu/kplrcasjobs/GOHelpKC.aspx},  
and found 2614 candidate host stars (hosting 3390 planets) in common and
 327 stars (hosting 768 planets) among the \kepler\ Confirmed Planets hosts. 
The GCK catalog should enable investigation of the UV excess as a function of stellar
age, rotation, and metallicity, identification of UV-bright
(potentially young) stars, and provide UV photometry for other
astrophysically interesting systems in the \kepler\ field.


\section{Description of the catalog file}                            
\label{sec:file}

Table~\ref{table:keywords} provides the description of the fields in
the GCK catalog file (columns 1 to 19).
The first keyword is the {\it gck\_id}, the main identifier of the GCK
catalog, with a sexagesimal, equatorial position-based source name
(i.e. GCK Jhhmmss.ss$+$ddmmss.s). The second and third keywords are
the coordinates of the NUV detection.
The following keywords give the photometry in magnitudes and fluxes,
with their corresponding errors.
The keyword {\it artifacts\_flags} provides the flags described in
Table \ref{table:artifactflags}.
For objects with a cross-match in the KIC the keywords {\it
ktswckey} and {\it kic\_keplerid}, are provided for the nearest
counterpart.  The keyword {\it ktswckey} is particularly useful to get
data from the table {\it keplerObjectSearchWithColors},
 also available in MAST through the tool CasJobs.
 This table contains the KIC catalog and other catalogs with coverage
 of the \kepler\ field. 
In Table~\ref{table:gcksample}, we show a portion of the GCK catalog.
The column tags designates the keyword number that also appear in
Table~\ref{table:keywords}. The GCK catalog is approximately 
$\sim$150 Mb (in ASCII format) and the table is available electronically 
with this paper.

As a useful reference for readers Table~\ref{table:gckwplanetssample} provides
  323 sources identified in the crossmatch between the GCK catalog and
  the Kepler targets with confirmed planetary  companions. A segment of this table is illustrated in 
 Table~\ref{table:gckwplanetssample} and contains the information
of columns 1-7, 10, 17, 20-28 listed in Table~\ref{table:keywords}.

\acknowledgements

The \kepler\ field observation was funded by Cornell University.
MO wishes to express his gratitude to the Astronomy Group of the
University of Rochester for their hospitality and to CONACyT for the
financial support received through the "Beca Mixta" program.
MC, EB, and MO also thank CONACyT for financial support through grants
SEP-2009-134985 and SEP-2011-169554.
EEM acknowledges support from NSF award AST-1313029.
We thank Chase Million for useful discussions.
\galex\ (Galaxy Evolution Explorer) is a NASA Small Explorer, launched
in 2003 April.
We gratefully acknowledge NASA's support for construction, operation,
and science analysis for the \galex\ mission, developed in cooperation
with the Centre National d'\'{E}tudes Spatiales of France and the
Korean Ministry of Science and Technology.

\bibliography{ms} 

 \clearpage

 \begin{landscape}
\begin{deluxetable*}{llll}[h!] 
\tabletypesize{\scriptsize}
\tablecaption{GCK catalog tags}
\tablewidth{0pt}
\tablehead{Tag Number & Tag Name & Description & Unit}
\startdata
1	&	gck\_id	&	\galex\ CAUSE Kepler (GCK) Identifier	&	number	\\
2      &      alpha\_j2000 &  Right ascension & Decimal degrees \\
3	&      delta\_j2000 &  Declination & Decimal degrees \\
4	&	nuv\_mag	&	Calibrated NUV magnitude	&	AB magnitude	\\
5	&	nuv\_magerr	&	Error of the calibrated NUV magnitude	&	AB magnitude	\\
6	&	nuv\_mag\_cor	&	Corrected calibrated NUV magnitude with Camarota \& Holberg (2014) calibration & AB magnitude	\\
7	&	nuv\_magerr\_cor	&	Error of the corrected calibrated NUV magnitude	&	AB magnitude	\\
8	&	nuv\_flux	&	NUV flux 	&	counts/sec	\\
9	&	nuv\_fluxerr	&	Error of NUV flux	&	counts/sec	\\
10	&	nuv\_s2n	&	Signal-to-noise ratio of NUV flux	&	dimensionless	\\
11	&	nuv\_bkgrnd\_mag	&	NUV background surface brightness at source position	&	AB magnitude	\\
12	&	nuv\_bkgrnd\_flux	&	Background NUV flux at source position	&	counts/sec/arcsec$^2$	\\
13	&	nuv\_exptime	&	Effective exposure time at source position	&	seconds	\\
14	&	fov\_radius	&	Distance of source from center of tile 	&	degrees	\\
15	&	artifact\_flags	&	Logical OR of artifact flags 	&	number	\\
16	&	ktswckey	&	Sequential number in CasJobs Kepler Colors Table of match &	number	\\
17	&	kic\_keplerid	&	Kepler Input Catalog identifier of match	&	number	\\
18	&	scan	&	GCK scan number	&	dimensionless	\\
19	&	image	&	GCK image number 	&	dimensionless	\\
20    &      kep\_name & Kepler host star name & dimensionless	\\
21    &   	kic\_kepmag & Kepler-band magnitude & AB magnitude \\
22	&   	kic\_g & g-band Sloan magnitude from the Kepler Input Catalog & AB magnitude \\
23    &   	kic\_r & r-band Sloan magnitude from the Kepler Input Catalog & AB magnitude \\
24    &   	kic\_i & i-band Sloan magnitude from the Kepler Input Catalog & AB  magnitude \\
25	&   	kic\_z & z-band Sloan magnitude from the Kepler Input Catalog & AB magnitude \\
26    &   	twomass\_j & 2MASS J-band magnitude & Vega magnitude \\
27	&   	twomass\_h & 2MASS H-band magnitude & Vega magnitude \\
28	&   	twomass\_k & 2MASS K-band magnitude & Vega magnitude \\
\enddata
 \label{table:keywords}
\end{deluxetable*}

\begin{deluxetable*}{ccccccccccccccccccc}[h!] 
\tabletypesize{\scriptsize}
\tablecaption{GCK catalog sample}
\tablewidth{0pt}
\tablehead{&&&&&&&&Column &&&&&&&&&& \\ (1)&(2)&(3)&(4)&(5)&(6)&(7)&(8)&(9)&(10)&(11)&(12)&(13)&(14)&(15)&(16)&(17)&(18)&(19)}
\startdata
GCK J18461428+4420359& 281.55948& 44.34330& 23.768& 0.638&-999&-999& 0.033& 0.02& 1.7& 26.95& 0.002& 1013.0& 0.843&   32&792984&8344414&    2&    2\\
GCK J19531681+4926274& 298.32004& 49.44096& 20.854& 0.072&-999&-999& 0.490& 0.03& 15.1& 26.15& 0.004& 1262.5& 0.754&   32&None&None&   15&    4\\
GCK J19325850+4922011& 293.24375& 49.36697& 20.038& 0.055&-999&-999& 1.040& 0.05& 19.8& 26.60& 0.002& 692.6& 0.497&   17&None&None&   12&    5\\
GCK J19432870+3909313& 295.86958& 39.15869& 14.803& 0.003& 14.681&  0.220& 129.076& 0.35& 366.7& 25.79& 0.005& 1101.0& 0.661&   48&658168&4075067&    9&   14\\
GCK J19183283+3827439& 289.63681& 38.46219& 22.907& 0.285&-999&-999& 0.074& 0.02& 3.8& 26.34& 0.003& 1524.7& 0.596&    1&12145669&None&    4&   12\\
GCK J19145926+4453507& 288.74693& 44.89742& 22.893& 0.369&-999&-999& 0.075& 0.03& 2.9& 26.52& 0.003& 1372.3& 0.197&    4&4369546&8681571&    7&    7\\
GCK J19503417+4804597& 297.64238& 48.08325& 22.789& 0.251&-999&-999& 0.083& 0.02& 4.3& 26.01& 0.004& 1593.2& 0.625&   32&15256924&None&   14&    5\\
GCK J19290783+4543284& 292.28261& 45.72457& 20.826& 0.093&-999&-999& 0.503& 0.04& 11.7& 26.48& 0.003& 611.5& 0.657&   32&None&None&   10&    7\\
GCK J19550979+4151533& 298.79079& 41.86481& 19.239& 0.037&-999&-999& 2.170& 0.07& 29.3& 25.64& 0.006& 875.6& 0.739&   32&6369109&6469387&   12&   14\\
GCK J19523858+4422221& 298.16074& 44.37282& 16.702& 0.006&-999&-999& 22.440& 0.12& 193.2& 25.98& 0.004& 1819.6& 0.415&   20&14957236&None&13&    8\\
\enddata
\tablecomments{Column number correspond to tag number in
  Table~\ref{table:keywords}. A -999 value indicates a non available
  data.}
 \label{table:gcksample}
\end{deluxetable*}


\begin{deluxetable*}{cccccccccccccccccc}[h!] 
\tabletypesize{\scriptsize}
\tablecaption{NUV magnitudes and ancillary data for Kepler targets with confirmed exoplanets}
\tablewidth{0pt}
\tablehead{&&&&&&&&Column &&&&&&&&& \\ (17)&(20)&(1)&(2)&(3)&(4)&(5)&(6)&(7)&(10)&(21)&(22)&(23)&(24)&(25)&(26)&(27)&(28)}
\startdata
 8359498 & Kepler-77&GCK J19182585+4420438& 289.60772& 44.34550& 20.575& 0.056&-999&-999& 19.43& 13.938& 14.449& 13.871& 13.720& 13.658& 12.757& 12.444& 12.361\\
8644288 &Kepler-18&GCK J19521905+4444474& 298.07939& 44.74650& 20.842& 0.062&-999&-999& 17.62& 13.549& 14.160& 13.479& 13.287& 13.187& 12.189& 11.872& 11.756 \\
6616218 & Kepler-314&GCK J19384180+4204320& 294.67418& 42.07566& 19.777& 0.035&-999&-999& 31.03& 12.557& 13.126& 12.459& 12.313& 12.205& 11.242& 10.849& 10.778 \\
9821454 & Kepler-59&GCK J19080950+4638249& 287.03957& 46.64025& 19.263& 0.026&-999&-999& 41.77&  14.307 &  14.669 & 14.259 & 14.152 & 14.116 & 13.253 & 12.974 & 12.928  \\
9595827 & Kepler-71&GCK J19392762+4617097& 294.86508& 46.28602& 21.969& 0.226&-999&-999& 4.80& 15.127& 15.692& 15.061& 14.885& 14.803& 13.926& 13.550& 13.468 \\
9884104 & Kepler-219&GCK J19145734+4645455& 288.73890& 46.76264& 19.609& 0.031&-999&-999& 35.27& 13.764& 14.174& 13.692& 13.588& 13.537& 12.678& 12.400& 12.388 \\
11121752 & Kepler-380&GCK J18493471+4845327& 282.39464& 48.75907& 18.278& 0.031&-999&-999& 35.46& 13.652& 14.047& 13.614& 13.483& 13.464& 12.614& 12.349& 12.288 \\
2302548 & Kepler-261&GCK J19252754+3736322& 291.36477& 37.60894& 21.181& 0.143&-999&-999& 7.58& 13.562& 14.271& 13.495& 13.259& 13.118& 12.127& 11.672& 11.585 \\
8572936 & Kepler-34&GCK J19454460+4438294& 296.43583& 44.64151& 20.884& 0.089&-999&-999& 12.25& 14.875& 15.372& 14.830& 14.662& 14.575& 13.605& 13.301& 13.237 \\
6850504 & Kepler-20&GCK J19104751+4220188& 287.69795& 42.33856& 18.900& 0.019&-999&-999& 56.16& 12.498& 12.997& 12.423& 12.284& 12.209& 11.252& 10.910& 10.871 \\
\enddata
\tablecomments{Column number correspond to tag number in
  Table~\ref{table:keywords}. A -999 value indicates an unavailable datum. Table~\ref{table:gckwplanetssample} is available in its entirety online as a 
  machine readable table.}
 \label{table:gckwplanetssample}
\end{deluxetable*}


\clearpage
\end{landscape}

\end{document}